www.nature.com/scientificreports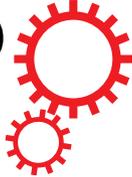

# Comparative study of microwave radiation-induced magnetoresistive oscillations induced by circularly- and linearly-polarized photo-excitation

Received: 08 April 2015
Accepted: 10 September 2015
Published: 09 October 2015Tianyu Ye[1], Han-Chun Liu[1], Zhuo Wang[1], W. Wegscheider[2] & Ramesh G. Mani[1]A comparative study of the radiation-induced magnetoresistance oscillations in the high mobility GaAs/AlGaAs heterostructure two dimensional electron system (2DES) under linearly- and circularly-polarized microwave excitation indicates a profound difference in the response observed upon rotating the microwave launcher for the two cases, although circularly polarized microwave radiation induced magnetoresistance oscillations observed at low magnetic fields are similar to the oscillations observed with linearly polarized radiation. For the linearly polarized radiation, the magnetoresistive response is a strong sinusoidal function of the launcher rotation (or linear polarization) angle, $\theta$. For circularly polarized radiation, the oscillatory magnetoresistive response is hardly sensitive to $\theta$.Microwave radiation induced zero resistance states[1] and associated microwave radiation-induced magnetoresistance oscillations[1,2] are thought to convey a novel steady-state non-equilbrium condition of the photo-excited high mobility 2D electron system. At low temperatures, in a perpendicular magnetic field, and under microwave photo-excitation, the magnetoresistance in the high mobility 2DES exhibits giant, periodic-in-$B^{-1}$, 1/4-cycle shifted magneto-resistance oscillations[1]. At lower temperatures, under moderate microwave power, the oscillatory minima turn into zero resistance states. Interesting experimental features revealed by prior studies in this field include: (a) the 1/4-cycle phase shift ref. 3,4, (b) the non-linear increase in the amplitude of the radiation-induced oscillations with the microwave power[5], (c) the sinusoidal dependence of the oscillation amplitude on the linear polarization angle[6,7], (d) observed correlations between the magneto-resistance oscillations and the microwave reflection from 2DES[8,9], and other fascinating phenomena[10–41].

These observed experimental phenomena have been considered in light of three principal theories for the photo-excited transport in the 2DES. Here, the displacement model[42–44] describes impurity and phonon scattering in the presence of inter- or intra- Landau Level microwave excitation, which regularly enhances or suppresses the back-scattering of electrons[45]. The microwave driven electron orbital model[46,47], which follows the periodic motion of the electron orbit centers under irradiation, has been particularly successful in explaining the polarization and power dependence of the radiation-induced magnetoresistance oscillations. The inelastic model[48], which explores the effects of a radiation-induced steady state non-equilibrium distribution, has proposed a radiation-induced oscillatory photoconductivity that is insensitive to the microwave polarization and grows linearly with the microwave power. There exist also a number of other interesting theories; the reader is encouraged to read this literature[49–56].

[1]Department of Physics and Astronomy, Georgia State University, Atlanta, Georgia 30303, USA. [2]Laboratorium für Festkörperphysik, ETH-Zürich, 8093 Zürich, Switzerland. Correspondence and requests for materials should be addressed to T.Y. (email: tye2@student.gsu.edu)SCIENTIFIC REPORTS | 5:14880 | DOI: 10.1038/srep148801

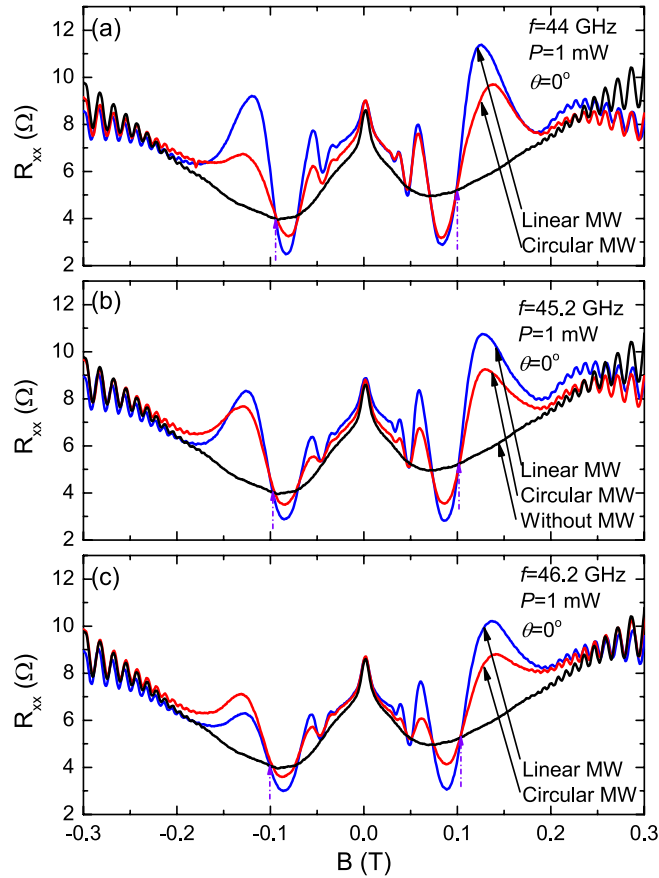

**Figure 1.** (**a**) Magnetoresistance measurements under microwave photo-excitation for a GaAs/AlGaAs 2DES specimen at (**a**) $f = 44$ GHz with $P = 1$ mW, (**b**) $f = 45.2$ GHz with $P = 1$ mW and (**c**) $f = 46.2$ GHz with $P = 1$ mW. Blue curves represent measurements with linearly polarized microwaves, red curves represent measurements with circularly polarized microwaves, and black curve represents measurements without microwave excitation. Magenta arrows pointing up indicate the magnetic fields for cyclotron resonance at each microwave frequency.

The role of the polarization angle in experiments that utilize linearly polarized microwave photo-excitation has been a topic of interest in recent work[6,7,57,58]. Associated experiments have shown, remarkably, that the amplitude of the radiation-induced magnetoresistance oscillations varies sinusoidally with the linear polarization angle, following a cosine-square function[7]. So far as circularly polarized microwave photo-excitation is concerned, an experimental study[19] examining the magnetotransport response for circular polarization reported on the immunity of the radiation-induced magnetoresistance oscillations to the polarization orientation for both circularly polarized and linearly polarized radiation. On the theoretical side, Lei and Liu examined radiation-induced magnetoresistance oscillations under a variety of polarization conditions, including linearly polarized microwaves with different polarization directions, and circularly polarized microwaves with left handed and right handed orientations[45,59]. They found that the amplitude of the magnetoresistance oscillations differs with the type of polarization of the radiation.

We have carried out a systematic comparative study of radiation-induced magnetoresistance oscillations using circularly polarized- and linearly polarized- microwaves, measured in the same sample, in a single cooldown, under nearly the same experimental conditions. The results show a striking sensitivity in the amplitude of the radiation-induced magnetoresistance oscillations under launcher rotation for linearly polarized microwaves, which is absent in the similar experiment carried out with circularly polarized microwaves. In addition, nearly similar response is observed in the cyclotron resonance active- and inactive- conditions for the circularly polarized radiation at the examined frequencies.

## Results

Figure 1 compares the observed microwave induced magnetoresistance oscillations under linearly- and circularly-polarized microwave excitation at $f = 44$, 45.2, and 46.2 GHz. Here, similar $B^{-1}$-periodic and 1/4-cycle shifted magnetoresistance oscillations are observed for both linearly- and circularly- polarized radiation, as the peaks and valleys shift together in the same way on $B$ axis as radiation-frequency changes in both cases. As well, there are some perceptible differences in the amplitude of the magnetoresistive response for the two types of polarizations. At every frequency, the amplitude of the microwave induced oscillatory magnetoresistance for circularly polarized microwaves is generally smaller than the amplitude of the oscillatory magnetoresistance induced by linearly polarized microwaves. An exception here is the first peak observable at $B = -0.125$ Tesla at 46.2 GHz, see Fig. 1(c). Although the height of this resistance peak is higher for circularly polarized radiation than it is for linearly polarized radiation, the total amplitude of the corresponding oscillation is nearly the same for the two polarizations. Here, it should be noted that although the source power is the same, i.e., $P = 1$ mW, for the measurements with linearly and circularly polarized radiation, the power at the sample will not be the same since additional hardware, in the form of the circular polarizer and horn, for the circular polarization measurements, introduces insertion loss and associated power attenuation.

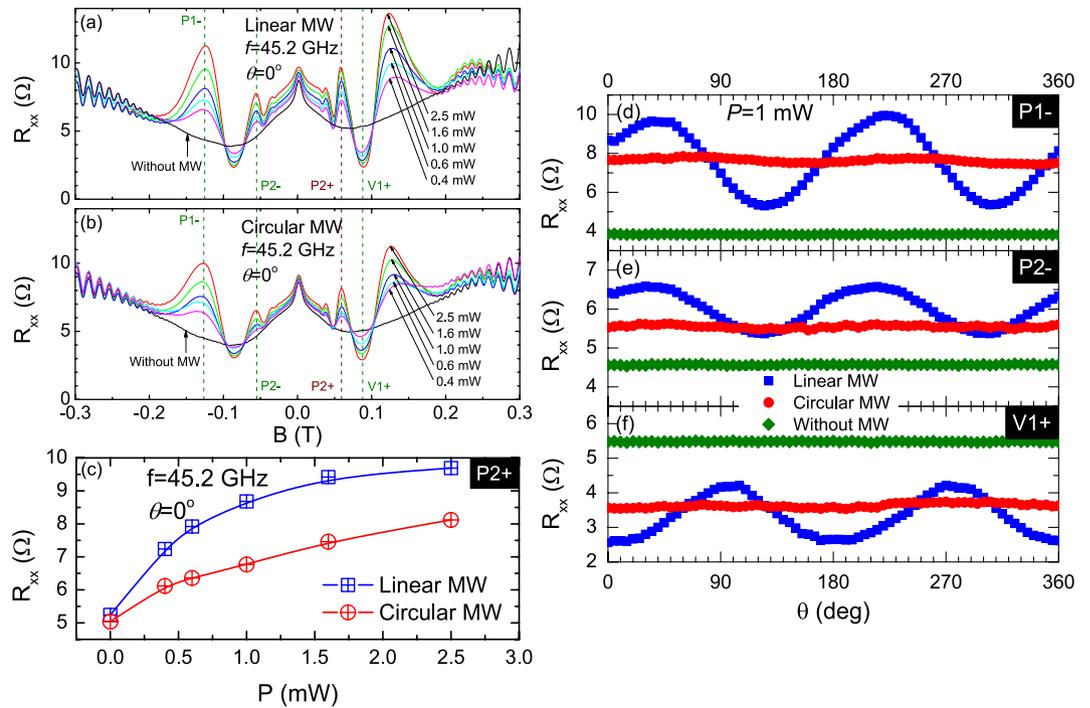

**Figure 2.** Magnetoresistance traces at different microwave power levels for (**a**) linearly polarized and (**b**) circularly polarized microwave radiation. (**c**) The diagonal resistance is shown as a function of the microwave power for linearly (blue) and circularly (red) polarized microwaves. These data were extracted from (**a**,**b**) at the magnetic field corresponding to P2+. Panels (**d**–**f**) show the diagonal resistance as a function of the antenna angle for $f = 45.2$ GHz radiation with source power $P = 1$ mW for both linearly polarized (blue symbols) and circularly polarized (red symbols) microwaves at (**d**) P1−, (**e**) P2− and (**f**) V1+. The responses in the absence of microwaves, i.e., dark, are shown in green. Here, the magnetic fields corresponding to P1-, P2- and V1+ are marked in (**a**,**b**).

Figure 2(a,b) show $R_{xx}$ vs. $B$ at a number of power levels, $P$, for both linearly and circularly polarized microwaves at 45.2 GHz. Here, as the microwave power increases, the oscillatory amplitude increases for both linearly- and circularly- polarized radiation. The power dependence of the peak resistance labeled P2+, shown in Fig. 2(c), indicates a non-linear increase in the peak resistance with power in the two cases. Note that, in Fig. 2(c), the $R_{xx}$ vs $P$ traces for the linearly- and circularly- polarized radiation start similarly but then diverge at higher source power. This feature indicates that less excitation is being coupled into the physical mechanism responsible for the radiation-induced magnetoresistance oscillations in the case of circularly polarized radiation.

Figure 2(d–f) shows the launcher-angle-dependence of the radiation-induced magnetoresistance oscillations for the linearly- and circularly-polarized radiations at the magnetic fields labeled as P1−, P2− and V1+, indicated as dashed lines in Fig. 2(a,b). Figure 2(d–f) show that for linearly polarized radiation, the magnetoresistive response at the oscillatory peaks and valley vary sinusoidally with the launcher angle as reported previously[7]. In sharp contrast, for circularly polarized radiation, the magnetoresistive response at the peaks and valley hardly show any launcher angle dependence; there appears to be a nearly constant response as a function of launcher angle. Finally, at a given peak or valley of the radiation-induced oscillatory magnetoresistance, the data for the circular polarization lie between the peak and valley of the oscillatory data for the linearly polarized radiation.

## Discussion

This study has compared the magnetoresistive response of the high mobility GaAs/AlGaAs 2D electron system under linearly- and circularly- polarized microwave photo-excitation in the frequency band spanning $43 \leq f \leq 50$ GHz. The results show that (a) peaks and valleys in the radiation-induced oscillatory magnetoresistance shift to higher magnetic field with an increase in the microwave frequency for circularly polarized radiation, similar to the response for linearly polarized radiation. (b) The amplitude of the radiation-induced oscillatory magnetoresistance increases non-linearly with the microwave power for circularly polarized radiation, similar to the observed response for linearly polarized microwaves. (c) Upon rotating the microwave launcher, the amplitude of the radiation-induced magnetoresistance oscillations hardly changes with the launcher angle for circularly polarized radiation. Indeed, nearly the same oscillatory magnetoresistive response is observed at all launch angles for circularly polarized radiation. This feature is quite unlike the strong sinusoidal variation in the response with the launch angle observed for linearly polarized radiation.

Among existing theoretical models, the displacement model of Lei and Liu has examined in detail the effect of different types of microwave polarization. Generally, this model[44,45,57,59] takes into account the Faraday configuration in experiment, where the magnetic field is perpendicular to the microwave electric field, which causes the electron to experience an additional Lorentz force due to the microwave field. With this approach, they have considered photo-excitation with an ac electric field $E = E_s \sin(\omega t) + E_c \cos(\omega t)$, which can serve to represent both circularly- and linearly- polarized microwaves. The carrier transport is then described by a drift velocity that consists of a dc part and an ac part following the microwave electric field as above, an electron temperature satisfying the requisite force and energy balance equations[59], with the frictional force including impurity and phonon scattering, and energy balance weighing the energy absorption rate from the radiation field against the energy dissipation rate from the electron system to the lattice. The model simulation from this theory for linearly polarized radiation[57] indicates a sinusoidal magnetoresistance change vs. the polarization angle, qualitatively similar to the results observed in experiment, see Fig. 3(d–f). The model simulation for circularly polarized radiation[45,59] produces magnetoresistance oscillations vs. the magnetic field that are qualitatively similar to the microwave-induced magnetoresistance oscillations observed for linearly polarized radiation, as shown here in Fig. 1. In addition, however, the theory indicates a difference in the amplitude of the magnetoresistive response for circular polarized radiation corresponding to the cyclotron-resonance active and inactive conditions, respectively, and also a difference in response for circularly- and linearly-polarized radiation, respectively. Although a larger magnetoresistive response is expected for the cyclotron resonance active condition in this model, it is worth noting that the magnetoresistance oscillations do not vanish completely in the cyclotron resonance inactive condition[59]. Indeed, calculations suggest only a factor of two change in the amplitude of the magnetoresistance oscillations for the cyclotron resonance active and inactive conditions at $f = 100$ GHz, see Fig. 1 in ref. 59. In the present experiment, the right hand circularly polarized microwaves correspond to the cyclotron resonance inactive condition at positive magnetic field because the polarization direction is against the cyclic motion of electrons. On the other hand, at negative magnetic field, the right hand polarized microwaves correspond to the cyclotron resonance active condition. Thus, the theory[44,45,57,59] suggests a possible greater response at negative magnetic fields in our experiments. Our results for $43 \leq f \leq 50$ GHz show, however, that the magnetoresistive oscillatory amplitudes for circularly polarized radiation are comparable for positive and negative magnetic fields. Since the theory presents numerically calculated diagonal resistances at specific frequencies, it is not possible at the moment to make a detailed comparison between experiment and theory, and determine possible sensitivity of theoretical expectations to parameters in the calculation, which could be a cause for the discrepancy. At a more qualitative level, it appears worth pointing out that traditional cyclotron resonance exhibits a strong sensitivity to the orientation of circular polarization because, when the circular polarization orientation matches the direction of cyclic motion of carriers, energy can be efficiently coupled from the radiation field to the charge carriers. Naively, one might expect also such a strong sensitivity of the radiation-induced magnetoresistance oscillations to the orientation of circular polarization of the electromagnetic radiation in this case. However, in the case of these radiation-induced magnetoresistance oscillations, there is the well-documented "1/4 cycle" phase shift of the magnetoresistance oscillations in inverse magnetic field with respect to the cyclotron resonance condition[3], which indicates nodes in the oscillatory magnetoresistance at $\omega/\omega_c \approx n$, with $n = 1, 2, 3…$. This feature suggests that the cyclotron resonance active- or inactive- conditions which play such a strong role in traditional cyclotron resonance, may not have such an overwhelming effect in these experiments examining the oscillatory magnetoresistance.

The microwave driven electron orbital model[58] has considered microwave electric fields that satisfy $\vec{E}(t) = \left(E_{0x}\vec{i} + E_{0y}\vec{j}\right)\cos(\omega t)$, which correspond to linearly polarized microwave radiation with a polarization angle $\alpha$ given by the relation $\tan \alpha = \frac{E_{0y}}{E_{0x}}$. This model successfully predicted the linear polarization angle dependence of MRIMOs. As well, it has been successful in describing the power dependence[60] with linearly polarized microwave radiation. It would be interesting for this model to examine in depth the model expectations for circularly polarized microwave radiation.

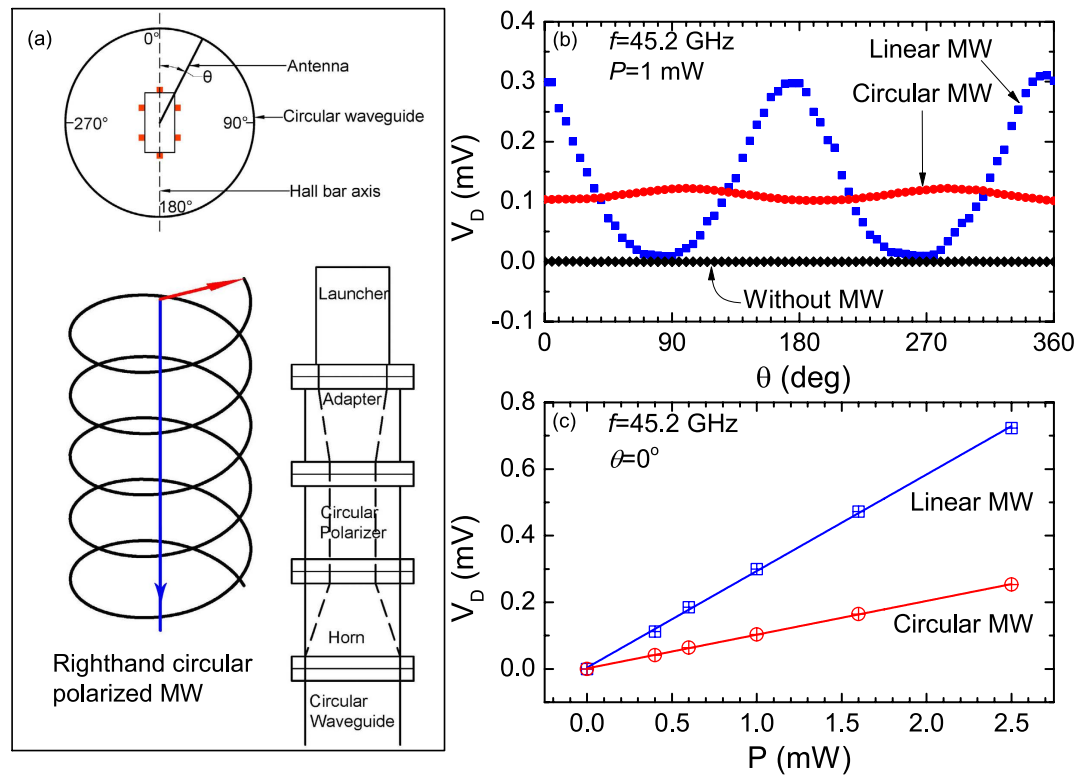

**Figure 3.** (**a**) The experimental setup for producing- and conveying- circularly polarized microwaves. Top: The Hall bar sample is centered with respect to a circular waveguide. The electric-dipole antenna that generates linearly polarized microwave radiation can be rotated with respect to the Hall bar axis. Bottom-right: The scheme for converting launcher-generated linearly polarized microwaves into circularly polarized microwaves using a commercially available adapter, circular polarizer, and microwave horn. This adapter, polarizer, and horn assembly are inserted for the circular polarization measurements. Bottom-left: This figure indicates the electric field and the propagation direction of right-handed circularly polarized microwaves. Red arrow indicates the direction of electric field and the blue arrow indicates the direction of transmission, which is parallel to the axis of the circular waveguide. (**b**) The observed microwave power detector response as a function of the launcher-antenna-angle when a microwave power detector is placed at the bottom of the sample holder, at the nominal position of the sample, for circularly- and linearly- polarized microwave radiation. Here, the measurements were carried out with 45.2 GHz microwave radiation at a power level $P = 1$ mW. The results for linearly polarized microwaves are represented with blue symbols while the results for circularly polarized microwaves are represented with red symbols. The black trace shows the power detector response in the absence of microwave radiation. (**c**) Microwave power detector response as a function of source microwave power for linearly (blue) and circularly (red) polarized microwave at 0° polarization angle.

To our knowledge, the inelastic model has not examined the linear- vis-a-vis circular- polarization issue at great depth since the theory indicates an absence of polarization sensitivity in (and a linear power dependence of) the radiation-induced magnetoresistance oscillations.

In summary, a comparative study of the radiation-induced magnetoresistance oscillations under linearly- and circularly- polarized microwave excitation indicates similar basic features in the observed oscillatory magnetoresistive response such as periodicity in $B^{-1}$, a 1/4-cycle phase shift, and non-linear increase in the amplitude of the oscillations with the microwave power for the two types of radiations. There is, however, a profound difference in the response observed upon rotating the microwave launcher for the linearly- and circularly- polarized radiations. For the linearly polarized radiation, the magnetoresistive response is a strong sinusoidal function of the launcher rotation (or linear polarization) angle, $\theta$. On the other hand, for circularly polarized radiation, the oscillatory magnetoresistive response is hardly sensitive to $\theta$. Finally, for circular polarized radiation, the magnetoresistive response for the cyclotron resonance active and inactive conditions is approximately the same over the entire field range. There could be a small difference in the immediate vicinity of cyclotron resonance. This feature is, at present, the topic of a more detailed investigation.

## Methods

Experiments were carried out on photolithographically fabricated Hall bars from molecular-beam-epitaxy grown high mobility GaAs/AlGaAs heterojunctions. A long cylindrical waveguide sample holder, with the Hall bar sample mounted at the bottom end, was inserted into a variable temperature insert (VTI), inside the bore of a superconducting solenoidal magnet. A sample temperature of 1.5 K was realized by pumping on- and reducing the vapor pressure of- the liquid helium within the VTI insert. As usual, the specimens inside the VTI reached the high mobility condition after brief illumination with a red light-emitting-diode.

Standard microwave components and techniques were utiled to realize- and convey-, linearly- and circularly- polarized microwave radiation to the specimens. A commercially available Agilent 83650B microwave synthesizer generated microwave excitation, which was conveyed by microwave coaxial cable to a rotatable microwave launcher at the top of the sample holder. The microwave launcher produces linearly polarized radiation and rotation of the launcher allows rotation of the polarization of the linearly polarized radiation. For the linearly polarized microwave experiment, the microwave launcher was connected to the circular waveguide sample holder via a rectangular- to circular- waveguide adapter. This adapter merely provides a smooth transition from rectangular to circular waveguide, and does not destroy the linear polarization of the microwave excitation. For the circularly polarized microwave experiments, a commercially available Millitech POL series circular polarizer, along with a microwave horn to provide a smooth transition, was inserted between the rectangular- to circular- waveguide adapter and the circular waveguide, see Fig. 3(a). This circular polarizer converts the linearly polarized microwave generated by the launcher to right-hand circularly polarized microwave in the frequency band between 43 GHz and 50 GHz.

In order to determine whether the microwave circular polarizer functioned as expected, a microwave power detector was installed at the nominal sample position, and the launcher angle was changed by rotating the launcher, above the circular waveguide, see Fig. 3(a). Here, the launcher angle is defined as the angle between the antenna within the launcher and a reference mark, which corresponds to $\theta = 0$, on the circular waveguide. Figure 3(b) exhibits the detected power vs. the launcher angle for 45.2 GHz microwave radiation, as measured by the microwave power detector. Here, the black trace shows the detected microwave power signal in the absence of microwaves. Note that the detected power vanishes for all $\theta$ in the absence of microwave power, as expected.

The blue curve shows the detected microwave power with launcher rotation angle $\theta$ for linearly polarized microwaves. This microwave power signal exhibits a sinusoidal variation with the launcher angle and the highest detected power occurs at around $\theta = 0$, which indicates that maximum response (or coupling between source and detector) is realized when the launcher-antenna is parallel to the detector-antenna, as expected.

The red curve exhibits the detected microwave power vs. launcher angle $\theta$ for circularly polarized microwaves. The red trace shows that the detected power for circularly polarized microwaves is insensitive to the launcher angle and the detected power in this instance is 2/3 of the average detected power for linearly polarized microwaves. Since circularly polarized radiation can be decomposed into two orthogonal, linearly-polarized components with a 1/4 wavelength phase shift, the circularly polarized radiation should present a constant reduced intensity to the detector under rotation of the microwave launcher, consistent with the measurement. The small angular dependence is attributed to the axial ratio rating (1 dB) of the linear - circular polarizer.

At $\theta = 0$, the detected power signal was also measured as a function of the microwave source-power, these results are shown in Fig. 3(c). From Fig. 3(c), it is clear that at the bottom end of circular waveguide sample holder, the detected microwave power changes linearly with the microwave source power, for both linearly polarized- and circularly polarized- microwave radiation. After these preliminary detected-power-measurements, the power detector was replaced by the Hall bar sample, with the long axis of the Hall bar aligned along the $\theta = 0$ direction.

Unlike the microwave type setup utilized here, the polarization dependence measurements reported in ref. 19. were carried out in a quasi-optical setup. The difference in setups can be attributed in part to the difference in frequency bands $100 \leq f \leq 350$ GHz band[19] vis-a-vis $43 \leq f \leq 50$ GHz band examined in the two studies. More specifically, the studies of Smet *et al.* utilized $4 \times 4$ mm$^2$ samples in the van-der-Pauw geometry mounted in an optical cryostat, with microwave excitation provided by backward wave oscillators over the range $100 \leq f \leq 350$ GHz. A series of wire grids, a mirror, a polarization transformer, attenuator, and lenses were utilized to direct the high frequency beam onto the specimen within the optical cryostat. The magnetotransport measurements, which showed radiation-induced magnetoresistance oscillations, helped to draw the conclusion that these oscillations are "entirely insensitive to the polarization state of the incident radiation" ref. 20.

## Acknowledgements
Magnetotransport measurements and T. Ye at Georgia State University are supported by the U.S. Department of Energy, Office of Basic Energy Sciences, Material Sciences and Engineering Division under DE-SC0001762. Additional activities and H-C. Liu are funded by the ARO under W911NF-14-2-0076.

## Author Contributions
T.Y. and R.M. conceived the experiment, provided the interpretation, and wrote the manuscript. Measurements by T.Y. and H.-C.L. W.W. provided the high quality MBE GaAs/AlGaAs wafers. Z.W. helped with the sample preparation.

## Additional Information
**Competing financial interests:** The authors declare no competing financial interests.

**How to cite this article**: Ye, T. *et al.* Comparative study of microwave radiation-induced magnetoresistive oscillations induced by circularly and linearly- polarized photo-excitation. *Sci. Rep.* **5,** 14880; doi: 10.1038/srep14880 (2015).